\documentstyle[epsf,12pt]{article}
\def\baselinestretch{1.1}
\title{\bf Time series model based on global structure of complete genome}
\author{Zu-Guo Yu$^{1,2}$ and Vo Anh$^{1}$\thanks{E-mail, Zu-Guo Yu: yuzg@hotmail.com or
 z.yu@qut.edu.au, Vo Anh: v.anh@qut.edu.au}\\
 {\small $^1$Centre for Statistical Science and Industrial Mathematics, Queensland University} \\
 {\small of Technology, GPO Box 2434, Brisbane, Q 4001, Australia.}\\
 {\small $^2$Department of Mathematics, Xiangtan University, Hunan 411105, P. R. China.\thanks{
 This is the permanent corresponding address of Zu-Guo Yu.}}
 }
\newcommand{\be}{\begin{equation}}
\newcommand{\ee}{\end{equation}}
\date{}
\begin{document}
\maketitle
\begin{abstract}
A time series model based on the global structure of the complete genome is proposed. Three kinds 
of length sequences of the complete genome are considered. The correlation dimensions and
Hurst exponents of the length sequences are calculated. Using these two exponents, some interesting
results related to the problem of classification and evolution relationship of bacteria are obtained.  
\end{abstract}
\vskip 0.2cm

{\bf PACS} numbers: 87.10+e, 47.53+n 

 {\bf Key words}:    Correlation dimension, Hurst exponent, 
Coding/noncoding segments, complete genome,

\section{Introduction}
 \ \ The nucleotide sequences stored in GenBank have exceeded hundreds of
millions of bases and they increase by ten times every five years. A great deal of
information  concerning origin of life,  evolution of species,  development 
of individuals, and  expression and regulation of genes, exist in these
sequences$^{\cite{luo}}$. In the past decade or so there has been an   
 enormous interest in unravelling the mysteries of DNA. It has become very important 
to improve on new theoretical methods to do DNA sequence analysis.
Statistical analysis of DNA sequences$^{[1-9]}$ using
modern statistical measures is proven to be particularly fruitful. 
There is another  approach to research DNA, 
 namely nonlinear scales method, such as fractal
dimension$^{\cite{luo4,luo2,juan,yhxc99}}$, complexity$^{\cite{YC,shen}}$. 
The correlation properties of coding
and noncoding DNA sequences was first studied by Stanley and coworkers$^{\cite{peng}}$
in their ``fractal landscape or DNA walk" model. The DNA walk defined in \cite{peng} 
is that
the walker steps ``up" if a pyrimidine ($C$ or $T$) occurs at position $i$ along the 
DNA chain, while the walker steps ``down" if a purine ($A$ or $G$) occurs at position $i$.
Stanley and coworkers$^{\cite{peng}}$ discovered there exists long-range correlation in
noncoding DNA sequences while the coding sequences correspond to regular random walk.
But if one considers more details by distinguishing $C$ from $T$ in pyrimidine, and $A$ from
 $G$ in purine (such as two or three dimensional DNA walk model$^{\cite{luo}}$ and
maps given in \cite{YC}), then  the presence of base correlation
has been found even in coding region.  However, DNA sequences are more
complicated than those these types of analysis can describe. Therefore, it is crucial
 to develop new tools for analysis with a view toward uncovering mechanisms
 used to code other types of information.  
 
 Since the
first complete genome of the free-living
bacterium {\it Mycoplasma genitalium} was sequenced in 1995$^{\cite{Fraser}}$,
 an ever-growing
number of complete genomes has been deposited in public databases.
The availability of complete genomes opens the possibility to
ask some global questions on these sequences.  The 
avoided and under-represented strings in some bacterial
complete genomes have been discussed in \cite{yhxc99,hlz98,hxyc99}.  
A time series model of CDS in complete genome has also been proposed in
 \cite{YW99}.

   One can  ignore the composition of the four kind of bases in coding and noncoding segments 
 and only consider the roughly structure of  the complete genome or long DNA sequences.  Provata 
 and  Almirantis
 $^{\cite{PY}}$ proposed a fractal Cantor pattern of DNA. They map coding segments to filled regions and 
 noncoding segments to empty regions of random Cantor set and then calculate the fractal dimension
 of the random fractal set. They found that the coding/noncoding partition in DNA sequences of lower organisms 
 is homogeneous-like, while in the higher eucariotes the partition is fractal. This result is interesting and
 reasonable, but it seems too rough to distinguish bacteria because the fractal dimensions of bacteria they gave
 out are all the same. The classification and evolution relationship of bacteria is one of the most important
 problem in DNA research. In this paper, we propose  a time series model based on the global structure of 
 the complete genome and we find that one can get more information from this model than that of the 
  fractal Cantor
 pattern. We have found some new results to the problem of classification and evolution relationship of
 bacteria.
    
\begin{figure}
\centerline{\epsfxsize=8cm \epsfbox{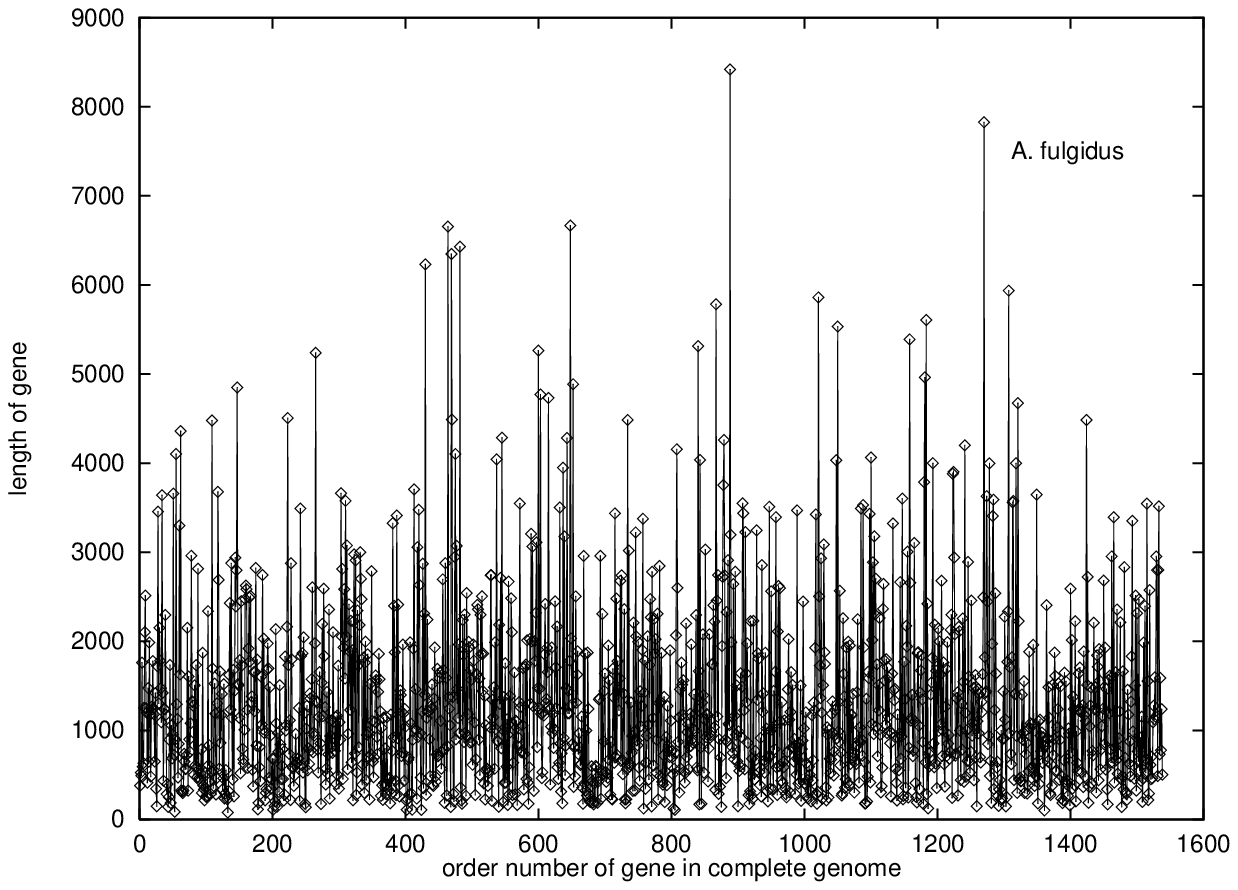}
  \epsfxsize=8cm \epsfbox{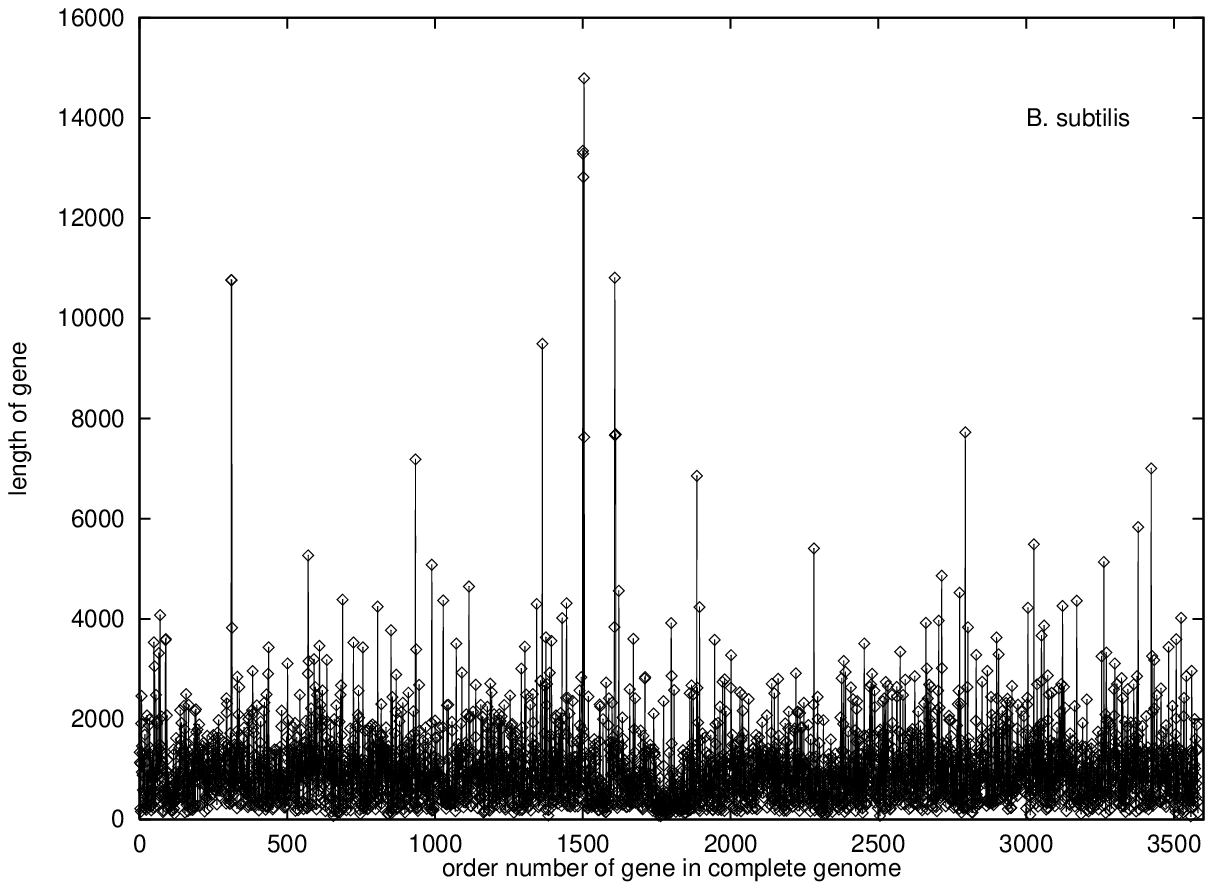}}
  \vskip 0.8cm
\centerline{ \epsfxsize=8cm \epsfbox{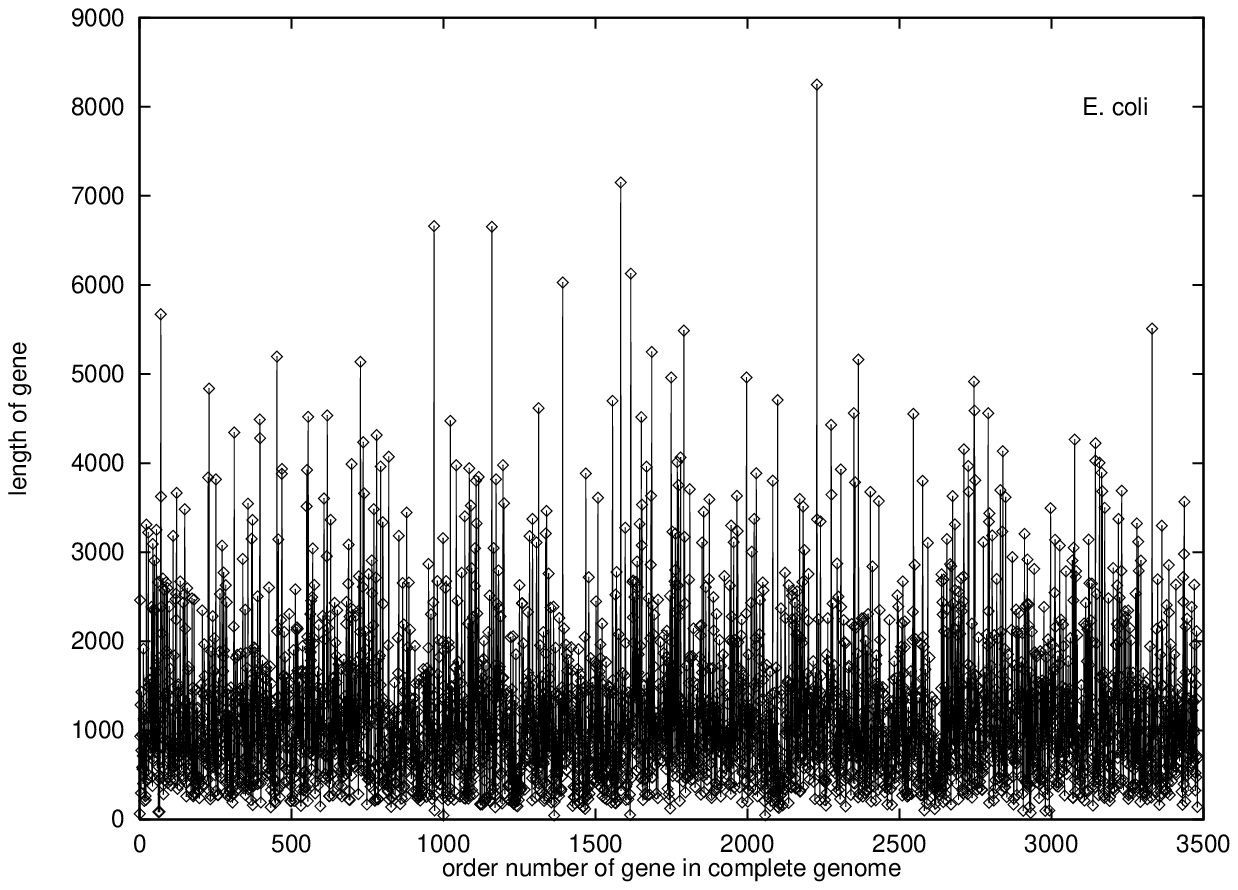}  
 \epsfxsize=8cm \epsfbox{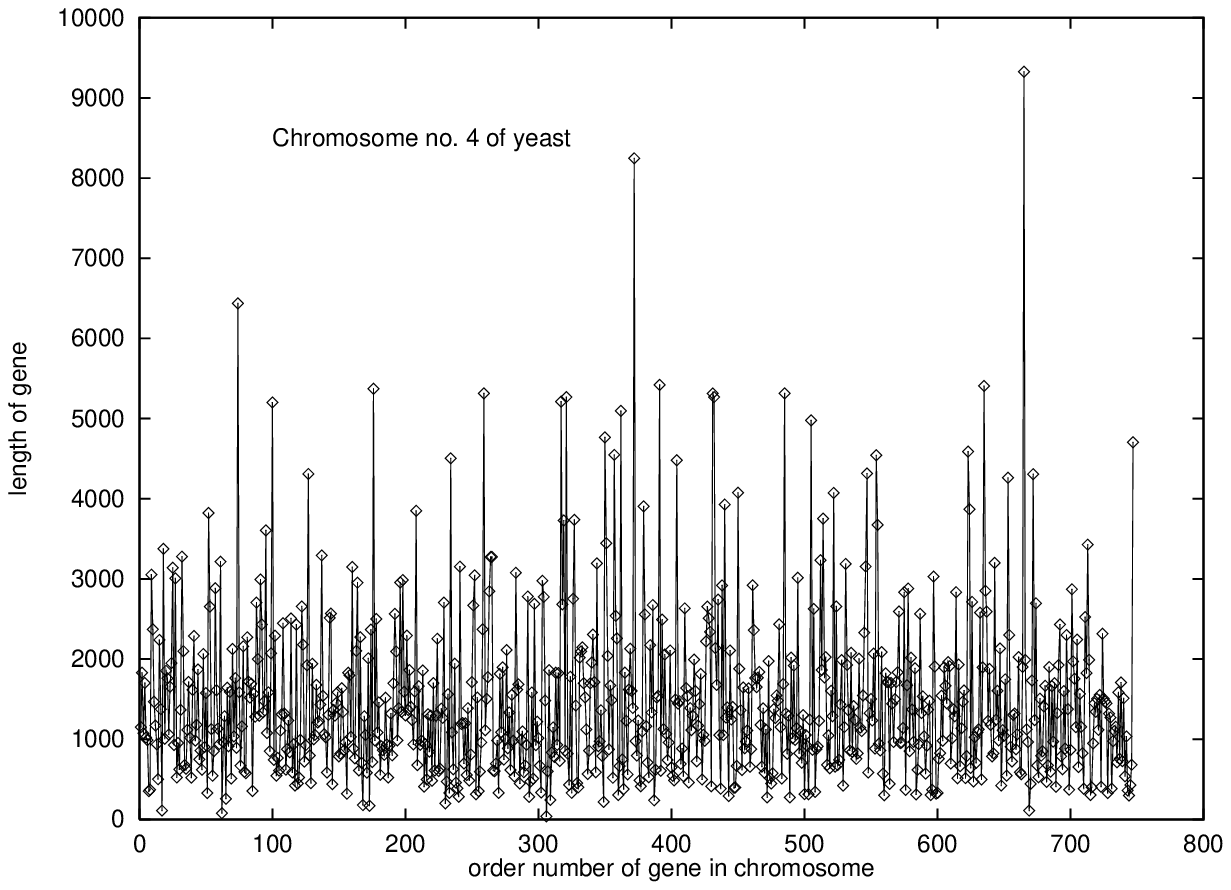} } 
\caption{Length and distribution of coding segments in the complete genome or Chromosome of some organisms. }
\label{distrib}
\end{figure}

    A DNA sequence is a sequence over
 the alphabet $\{A,C,G,T\}$ representing the four bases from which DNA is 
 assembled, namely adenine, cytosine, guanine, and thymine. But from views of the level of structure,
 the complete genome of organism is made up of coding and noncoding segments. Here
 the length of a coding/noncoding segment means the number of its bases. First we simply count out
 the lengths of coding/noncoding segments in the complete genome. Then we can get three kinds of integer sequences
 by the following ways.
    
    i)  First we order all lengths of coding and noncoding segments according to the
  order of coding and noncoding segments in the complete genome, then replace the lengths of noncoding
  segments by their negative numbers. So that we can distinguish lengths of coding and noncoding 
  segments. This integer sequence is named {\it whole length sequence}.
  
    ii)  We order all lengths of coding segments according to the order of
  coding segments in the complete genome. We name this integer sequence {\it coding length sequence}.
  For some examples, we plot the distribution of coding length sequences
  of three bacteria genome and the 4th chromosome of Saccharomyces 
  cerevisiae (yeast) in Figure \ref{distrib}.
  
  iii)  We order all lengths of noncoding segments according to the order of
  noncoding segments in the complete genome. This integer sequence is named
  {\it noncoding length sequence}.
  
     We can now view these three kinds of integer sequences as time series. We want to calculate 
     their  correlation
  dimensions and Hurst exponents.   
      
\section{Correlation dimension and Hurst exponent}
\begin{figure}
\centerline{\epsfxsize=8cm \epsfbox{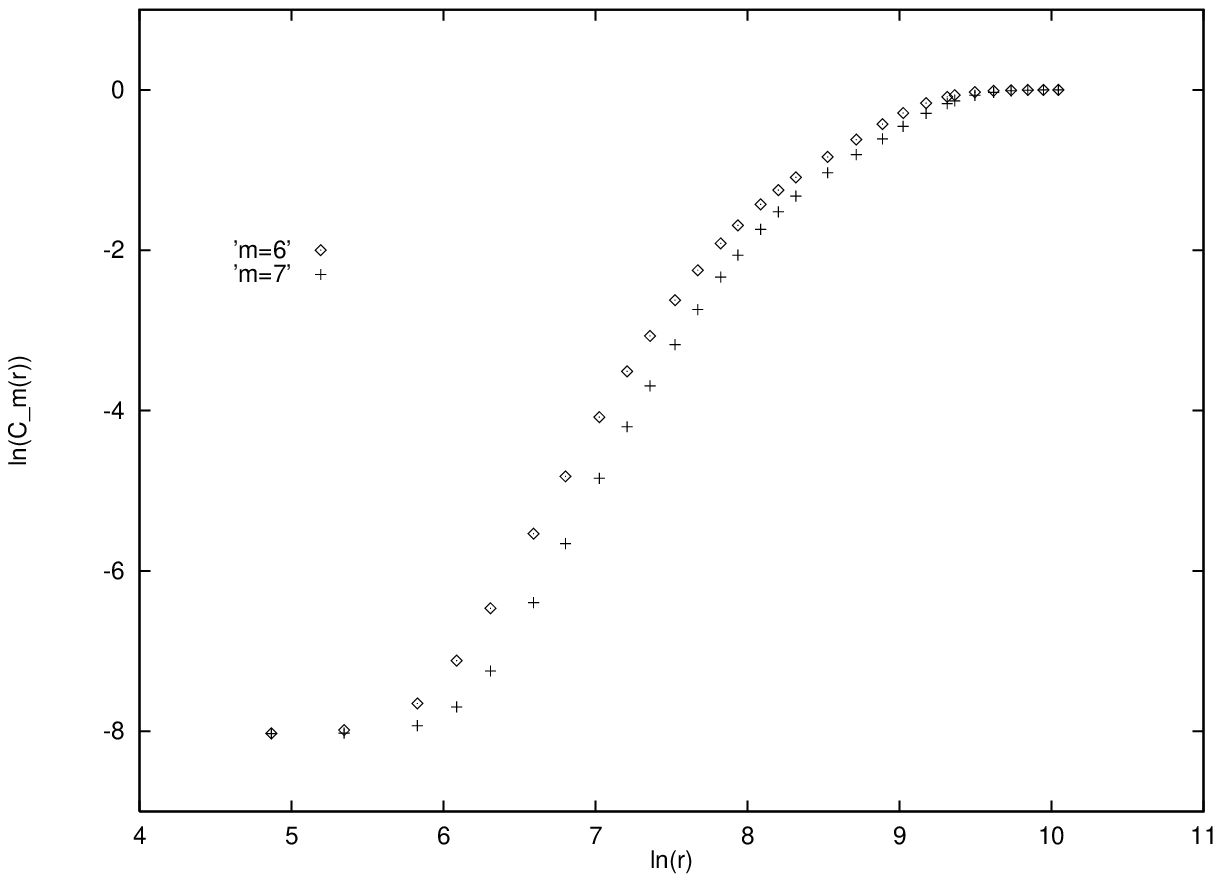}
  \epsfxsize=8cm \epsfbox{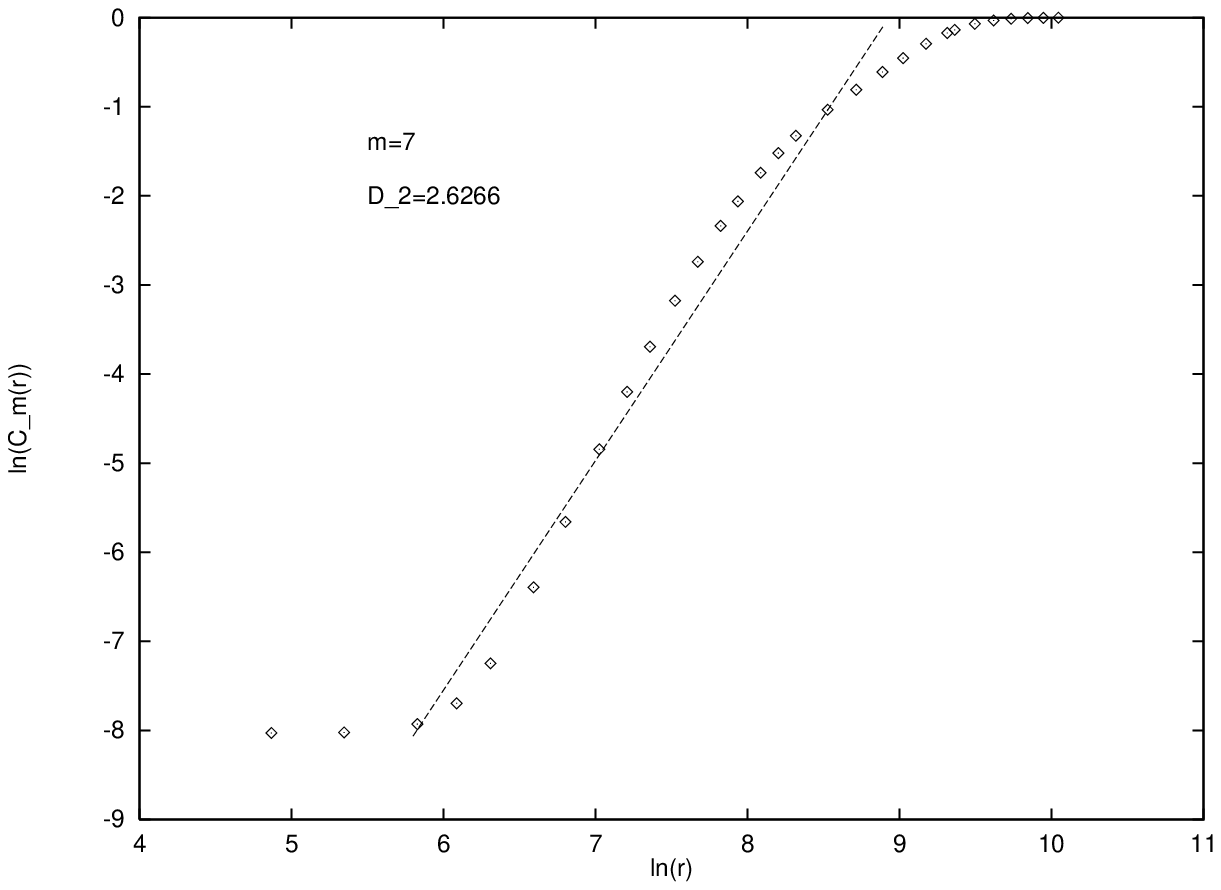}}
\caption{ Left) $\ln r$-$\ln C_m(r)$ figure of the length sequence of coding and 
noncoding segments of
A. fulgidus when m=6,7.  Right) Estimate of the correlation dimension (the continuous line).}
\label{D2fig}
\end{figure}

\ \ The notion of correlation dimension, introduced by Grassberger and
 Procaccia$^{\cite{GP1,GP2}}$,
suits well experimental situations, when only a single time series is available. It is now
being used widely in many branches of physical science.
 Consider a sequence of data from a computer or laboratory experiment:
\begin{equation}
x_1,\ x_2,\ x_3,\cdots,\ x_N,
\end{equation}
where $N$ is a large enough number. These numbers are usually sampled at an equal time interval
$\Delta\tau$. We embed the time series into ${\bf R}^m$, choose a time delay
 $\tau=p\Delta\tau$,
then obtain
\begin{equation}
{\bf y}_i=(x_i,x_{i+p},x_{i+2p},\cdots,x_{i+(m-1)p}),\ \ i=1,2,\cdots,N_m,
\end{equation}
where
\begin{equation}
N_m=N-(m-1)p.
\end{equation}
In this way we get $N_m$ vectors in the embedding space ${\bf R}^m$.

\ \ For any ${\bf y}_i,{\bf y}_j$, we define the distance as
\begin{equation}
r_{ij}=d({\bf y}_i,{\bf y}_j)=\sum_{l=0}^{m-1}|x_{i+lp}-x_{j+lp}|.
\end{equation}
If the distance is less than a given number $r$, we say that these two vectors are correlated.
The correlation integral is defined as
\begin{equation}
C_m(r)=\frac{1}{N_m^2}\sum_{i,j=1}^{N_m}H(r-r_{ij}),
\end{equation}
where $H$ is the Heaviside function
\begin{equation}
H(x)=\left\{\begin{array}{l}1,\quad {\rm if}\ x>0,\\
0,\quad {\rm if}\ x\le 0. \end{array}\right.
\end{equation}
For a proper choice of $m$ and not too big a value of $r$, it has been 
shown by Grassberger and
Procaccia$^{\cite{GP2}}$ that the correlation integral $C_m(r)$ behaves like
\begin{equation}
C_m(r)\ \propto\ r^{D_2(m)}.
\end{equation}
Thus one can define the correlation dimension as
\begin{equation}
D_2=\lim_{m\longrightarrow\infty}D_2(m)=\lim_{m\longrightarrow\infty}\lim_{r\longrightarrow 0}
\frac{\ln C_m(r)}{\ln r}.
\end{equation}
For more details on $D_2$, the reader can refer to \cite{Hao89}. 

\ \ To deal with practical problems, one usually choose $p=1$.  If
we choose a sequence $\{r_i:\ 1\le i\le n\}$ such that $r_1<r_2<r_3<\cdots <r_n$,
 then a scaling region can be found
in the $\ln r-\ln C_m(r)$ plane, see \cite{Hao89}, p.346.  Then the slop of the 
scaling region is $D_2(m)$. When $D_2(m)$ does not change with $m$ increasing, we can take
this $D_2(m_0)$ as the estimate value of $D_2$.  We calculate the correlation dimensions of
 three kinds of length sequences of the complete genome 
  using the method introduced above. From the $\ln r-\ln C_m(r)$ figures of 
 these sequences of different values of embedding
 dimension $m$, we find that it is suitable to choose $m=7$. For example, 
 we give the $\ln r-\ln C_m(r)$
 figure of whole length sequence of A. fulgidus when $m=6,7$ (Figure \ref{D2fig}). 
 We take the region from the third point to the 20th point
 (from left to right) as the scaling region.
    
Hurst$^{\cite{hurst}}$ invented  the now famous statistical method --- 
{\it the rescaled range analysis} ($R/S$ analysis) to study the long-range dependence 
in time series. Later on,  
B.~B.~Mandelbrot$^{\cite{mandelbrot}}$ and J. Feder $^{\cite{feder}}$  brought 
$R/S$ analysis into fractal analysis.
For any time series  $x=\{x_k\}_{k=1}^N$ and 
any $2\le n\le N$, one can define

\be <x>_{n}=\frac{1}{n}\sum_{i=1}^{n}x_i \ee

\be X(i,n)=\sum_{u=1}^i[x_u-<x>_{n}] \ee

\be R(n)=\max_{1\le i\le n}X(i,n)-\min_{1\le i\le n}X(i,n) \ee

\be S(n)=[\frac{1}{n}\sum_{i=1}^{n}(x_i-<x>_{n})^2]^{1/2}. \ee

Hurst found that

\be R(n)/S(n) \ \sim\ (\frac{n}{2})^H. \ee

 $H$ is called the {\it Hurst exponent}.

\begin{figure}
\centerline{\epsfxsize=12cm \epsfbox{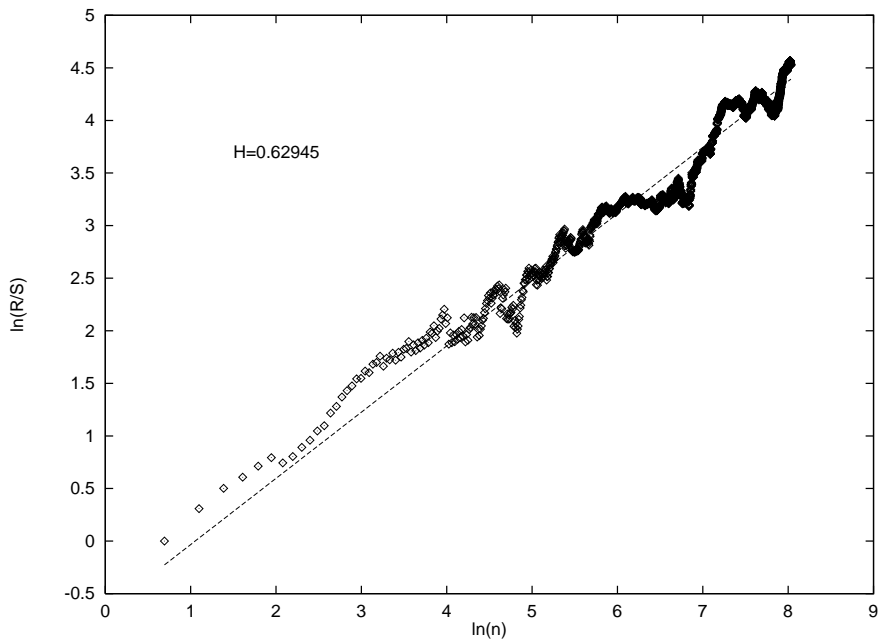}}
\caption{ Calculation of Hurst exponent.}
\label{hfig}
\end{figure}

  As $n$ changes from 2 to $N$, we obtain $N-1$ points 
in the $\ln(n)$ v.s. $\ln(R(n) / S(n))$
 plane. Then  we can calculate the Hurst exponent $H$ of the length sequence of organisms
using the 
least-squares linear fit. 
As an example, we plot the graph of $R/S$ analysis of the whole length sequence of 
A. fulgidus 
in Figure \ref{hfig}.

  The Hurst exponent is usually used as a measure of complexity. 
  The trajectory 
of the record is a curve with  fractal dimension $D=2-H$ (\cite{mandelbrot},p.149). Hence 
a smaller $H$ means a more complex system.
When applied to fractional Brownian motion,   
the system is said to be {\em  persistent} if $H > 1/2$, which means that if
for a given time period $t$, the motion is along one direction,
then in a succeeding time, it is more likely that the motion
will follow the same direction.  For $H < 1/2$,
the opposite holds, that is, the system is {\em antipersistent}. But when $H=1/2$,
the system is  a Brownian motion, and is random.

\section{ Data and results. }
\ \ \ \ More than 21  bacterial complete genomes are now available 
in public databases
. There are five Archaebacteria: Archaeoglobus fulgidus, 
Pyrococcus abyssi, Methanococcus jannaschii, Aeropyrum pernix 
and Methanobacterium thermoautotrophicum; four Gram-positive 
Eubacteria: Mycobacterium tuberculosis, Mycoplasma pneumoniae,
 Mycoplasma genitalium, and Bacillus subtilis. The others 
 are Gram-negative Eubacteria. These consist of  two Hyperthermophilic bacteria: Aquifex 
 aeolicus and Thermotoga maritima; 
 six proteobacteria: Rhizobium sp. NGR234, 
 Escherichia coli, Haemophilus influenzae,
  Helicobacter pylori J99, Helicobacter pylori 26695
  and Rockettsia prowazekii;  two 
 chlamydia Chlamydia trachomatis  and Chlamydia pneumoniae, 
 and two Spirochete: Borrelia burgdorferi  and Treponema pallidum.

 We calculate the correlation dimensions and Hurst exponents of three kinds of length sequences
 of the above 21 bacteria.  The estimated results are given in Table \ref{D2} 
 ( we denote by 
 $D_{2,whole}$, $D_{2,cod}$ and $D_{2,noncod}$ the correlation dimensions of whole, coding and 
 noncoding length sequences, from top to bottom, in the increasing order of  the value 
 of $D_{2,cod}$ ) and Table \ref{H}
  ( we denote by 
 $H_{whole}$, $H_{cod}$ and $H_{noncod}$ the Hurst exponents of whole, coding and 
 noncoding length sequences, from top to bottom, in the increasing order of the value 
 of $H_{cod}$ ).
 
  \begin{table}
\caption{$D_{2,whole}$, $D_{2,cod}$ and $D_{2,noncod}$ of 21  bacteria.}
\begin{center}
\begin{tabular}{|c|c|c|l|l|}
\hline\hline
  $D_{2,whole}$ & $D_{2,cod}$ &    $D_{2,noncod}$ & Species of Bacterium &\ \ \ \ Category\\
 \hline
 2.1126 & 1.3581 & 1.1612 & Mycoplasma genitalium & Gram-positive Eubacteria\\
 2.3552 & 1.7102 & 1.5077 & Mycoplasma pneumoniae & Gram-positive Eubacteria\\
 2.5239 & 1.8891 & 0.8944 & Aquifex aeolicus & Hyperthermophilic bacteria\\
 2.5125 & 1.9094 & 0.5849 & Thermotoga maritima & Hyperthermophilic bacteria\\
 2.2705 & 2.0119 & 2.2014 &  Rhizobium sp. NGR234 &Proteobacteria\\
 2.4060 & 2.0378 & 0.4695 & Borrelia burgdorferi & Spirochete\\
 2.4561 & 2.0729 & 0.6145 & Treponema pallidum  &  Spirochete\\
 2.5345 & 2.1674 & 1.3001 & Chlamydia trachomatis & Chlamydia \\
 2.6015 & 2.3055 & 1.3187 & Chlamydia pneumoniae & Chlamydia \\
 \hline\hline
 2.6096 & 2.4137 & 0.8475 & Pyrococcus abyssi & Archaebacteria\\
 2.5617 & 2.4589 & 2.1515 & Rickettsia prowazekii & Proteobacteria\\
 2.6266 & 2.4867 & 0.7011 & Archaeoglobus fulgidus & Archaebacteria\\
 2.6916 & 2.5195 & 1.2134 & Aeropyrum pernix & Archaebacteria\\
 2.6497 & 2.5248 & 0.9239 & Helicobacter pylori 26695 & Proteobacteria\\
 2.6353 & 2.5364 & 0.9555 & Helicobacter pylori J99 & Proteobacteria\\
 2.7181 & 2.8417 & 1.1126 & Haemophilus influenzae & Proteobacteria\\
 2.6558 & 2.8861 & 1.1427 & Methanococcus jannaschii & Archaebacteria\\
 2.5687 & 2.9097 & 0.6862 & M. thermoautotrophicum & Archaebacteria \\
 2.8473 & 2.9250 & 1.1031 & Mycobacterium tuberculosis & Gram-positive Eubacteria\\
 2.8984 & 3.0976 & 1.3660 & Escherichia coli &Proteobacteria\\
 2.7039 & 3.2435 & 1.1035 & Bacillus subtilis & Gram-positive Eubacteria\\
 \hline\hline
 \end{tabular}
 \end{center}
 \label{D2}     
\end{table}

 \begin{table}
\caption{$H_{whole}$,$H_{cod}$ and $H_{noncod}$ of 21  bacteria.}
\begin{center}
\begin{tabular}{|c|c|c|l|l|}
\hline\hline
    $H_{whole}$ & $H_{cod}$ & $H_{noncod}$   & Species of Bacterium &\ \ \ \ Category\\
 \hline
 0.3904 & 0.3311 & 0.6446 &  Rhizobium sp. NGR234 &Proteobacteria\\
0.4280 & 0.4108 & 0.5640 & Pyrococcus abyssi & Archaebacteria\\
 0.4063 & 0.4381 & 0.5925 & Rickettsia prowazekii & Proteobacteria\\
 0.4736 & 0.4660 & 0.5504 & Helicobacter pylori 26695 & Proteobacteria\\
 0.4828 & 0.5147 & 0.4648 & Mycoplasma genitalium & Gram-positive Eubacteria\\
 0.5064 & 0.5343 & 0.5381 & Chlamydia pneumoniae & Chlamydia \\
 0.5979 & 0.5365 & 0.5873 & Helicobacter pylori J99 & Proteobacteria\\
 0.4731 & 0.5445 & 0.6005 & Chlamydia trachomatis & Chlamydia \\
 0.5297 & 0.5698 & 0.5626 & Mycobacterium tuberculosis & Gram-positive Eubacteria\\
 0.5410 & 0.5882 & 0.4948 & Thermotoga maritima &  Hyperthermophilic bacteria\\
 0.5288 & 0.5941 & 0.6843 & Mycoplasma pneumoniae & Gram-positive Eubacteria\\
 0.5362 & 0.5985 & 0.4655 & Escherichia coli &Proteobacteria\\
\hline\hline
 0.5528 & 0.6017 & 0.3153 & M. thermoautotrophicum & Archaebacteria \\
 0.6295 & 0.6098 & 0.6311 & Archaeoglobus fulgidus & Archaebacteria\\
 0.6013 & 0.6145 & 0.4605 & Aquifex aeolicus & Hyperthermophilic bacteria\\
 0.5202 & 0.6153 & 0.5136 & Haemophilus influenzae & Proteobacteria\\
 0.5727 & 0.6371 & 0.4986 & Aeropyrum pernix &  Archaebacteria\\
 0.6830 & 0.6622 & 0.6764 & Borrelia burgdorferi & Spirochete\\
 0.7213 & 0.6894 & 0.5612 & Treponema pallidum & Spirochete\\
 0.7271 & 0.7183 & 0.6399 & Bacillus subtilis & Gram-positive Eubacteria\\
 0.7732 & 0.7793 & 0.3607 & Methanococcus jannaschii & Archaebacteria\\
 \hline\hline
 \end{tabular}
 \end{center}
 \label{H}     
\end{table}

\section{Discussion and conclusions}

\ \ \ \ Although the existence of the archaebacterial urkingdom has been accepted by many 
biologists, the classification of bacteria is still a matter of controversy$^{\cite{iwabe}}$.
The evolutionary relationship of the three primary kingdoms (i.e. archeabacteria, eubacteria
and eukaryote) is another crucial problem that remains unresolved$^{\cite{iwabe}}$. 

From Table \ref{D2}, we can roughly divide bacteria into two classes,    
  one class with $D_{2,cod}$ less than 2.40,  and the other with $D_{2,cod}$  greater
than 2.40.
We  observe that the classification of bacteria using $D_{2,cod}$  almost 
coincides with the traditional classification of bacteria. 
 All Archaebacteria  belong to the same class. All Proteobacteria belong to the same class except 
 Rhizobium sp. NGR234, in particular, the closest Proteobacteria Helicobacter pylori 26695
 and Helicobacter pylori J99  group with each other. Two Spirochete  group with each other.
 Two Chlamydia gather with each other. 
 Gram-positive bacteria  is divided into two sub-categories: Mycoplasma genitalium and 
  Mycoplasma pneumoniae belong to one class
 and gather with each other, Mycobacterium tuberculosis and Bacillus subtilis belong to 
 another class and almost gather with each other.
 
   If one classifies bacteria using $D_{2,whole}$, with the $D_{2,whole}$ of one 
   subclass  less than
   2.55, that of  the other larger than 2.55,  almost the same results hold as those 
   using $D_{2,cod}$.
   But when one classifies bacteria using $D_{2,noncod}$, the results are quite different. This is quite
   reasonable because the coding segments occupy the main part of space of the DNA chain of bacteria.

  A surprising feature shown in Table \ref{D2} is that the Hyperthermophilic bacteria 
  (including Aquifex aeolicus 
  and Thermotoga maritima) are 
linked closely with the Archaebacteria if we only consider the length sequences of 
noncoding segments. But when
we consider the length sequences of coding segments, they are linked closely with eubacteria.
We notice that Aquifex, like most 
Archaebacteria, is hyperthermophilic. Hence  it seems that their 
hyperthermophilicity property    
is possibly controlled by the noncoding part of the genome, contrary to 
the traditional view resulting from classification based on  the coding part
 of the genome. It has previously been shown that 
Aquifex has close relationship with Archaebacteria from the gene comparison of
 an enzyme needed for the synthesis of the amino acid 
 trytophan$^{\cite{pennisi}}$. 
 Such strong correlation on the level of complete 
 genome between Aquifex and Archaebacteria is not easily accounted for by lateral 
 transfer and other accidental events$^{\cite{pennisi}}$. Our result is based on 
 different levels of the genome from
 that used by the authors of \cite{pennisi}

   From Table \ref{D2}, one can also see the $D_{2,cod}$ values are almost larger than 
   the $D_{2,noncod}$ values. 
   Hence the coding
   length sequences are more complex than the noncoding length sequences.
   
   From Table \ref{H}, we can also roughly divide bacteria into two classes, 
     one class with $H_{cod}$ less than 0.60,  and the other with $H_{cod}$   greater
than 0.60. One can see all Archeabacteria belong to the same class except 
Pyrococcus abyssi.
All Gram-positive Eubacteria belong to the same class except Bacillus subtilis. 
All Proteobacteria belong 
to the same class except Haemophilus influenzae. Two Spirochete  group with each other.
 Two Chlamydia almost 
 group with each other.

    We also find the $H_{noncod}$ values of all Archeabacteria except Pyrococcus abyssi,
         two Hyperthermophilic bacteria, 
    and Mycoplasma genitalium and E. coli are less than $1/2$, while  those of other bacteria are greater than $1/2$.
  Hence  Hyperthermophilic bacteria have some common information with Archaebacteria in noncoding segments.  

  We calculate $D_{2,whole}$, $D_{2,cod}$, $D_{2,noncod}$ $H_{whole}$, $H_{cod}$ and $H_{noncod}$
  of the 4th chromosome of Saccharomyces 
  cerevisiae (yeast). 
  They are 
  2.5603, 2.1064, 2.5013, 0.5517, 0.6255 and 0.5482  respectively. From Tables \ref{D2} and \ref{H},
   if we consider $D_{2,whole}$, $H_{whole}$ and $H_{cod}$, 
  we can see that Archaebacteria
  and Chlamydia are linked more closely with yeast which belongs to eukaryote than 
  other categories of bacteria. There are several reports (such as  \cite{lhb}) that,
  in some RNA and protein species, archeabacteria are much more similar in sequences
  to eukaryotes than to eubacteria. Our present result supports this point of view. 

In  \cite{YC}, we find that the Hurst exponent is a good tool to distinguish
 different functional regions. But now   considering more global structure of the 
 genome, we find
 the correlation dimension  a better
 exponent to use for classification of bacteria than the Hurst exponent in this level.

\section*{ACKNOWLEDGEMENTS}
\ \ \   One of the authors Zu-Guo Yu would like to express his thanks to Prof. Bai-lin Hao of Institute of Theoretical
Physics of Chinese Academy of Science for introducing him into this field and continuous
 encouragement. He also wants to thank Dr. Bin Wang of ITP and Dr. Fawang Liu   at QUT for useful
 discussions about computer programmes. This project is supported by Postdoctoral Research
 Support Grant No. 9900658 of QUT.

\end{document}